\documentstyle[12pt]{article}

\newcommand{\eq}{\begin{equation}}
\newcommand{\en}{\end{equation}}
\newcommand{\bea}{\begin{eqnarray}}
\newcommand{\eea}{\end{eqnarray}}
\newcommand{\spz}{\hspace{0.7cm}}
\newcommand{\virg}{\spz,\spz}

\newcommand{\de}{\partial_u}
\newcommand{\half}{{\textstyle\frac{1}{2}}}
\newcommand{\sfrac}[2]{{\textstyle\frac{#1}{#2}}}

\newcommand{\bL}{\mbox{\bf L}}
\newcommand{\bM}{\mbox{\bf M}}

\newcommand{\bR}{\mbox{\bf R}}
\newcommand{\bT}{\mbox{\bf T}}

\newcommand{\buno}{\mbox{\bf 1}}

\newcommand{\cG}{{\cal G}}

\newcommand{\cL}{{\cal L}}

\newcommand{\cP}{{\cal P}}

\newcommand{\NP}[1]{Nucl.\ Phys.\ {\bf #1}}
\newcommand{\PL}[1]{Phys.\ Lett.\ {\bf #1}}

\newcommand{\CMP}[1]{Comm.\ Math.\ Phys.\ {\bf #1}}

\newcommand{\IJMP}[1]{Int.\ J.\ Mod.\ Phys.\ {\bf #1}}

\begin{document}
\sloppy
\renewcommand{\thefootnote}{\fnsymbol{footnote}}

\newpage
\setcounter{page}{1}

\vspace{0.7cm}
\begin{flushright}
DFUB 95-08\\
August 1995
\end{flushright}
\vspace*{1cm}
\begin{center}
{\bf GENERALIZED KdV AND QUANTUM INVERSE SCATTERING DESCRIPTION OF
  CONFORMAL MINIMAL MODELS}\\
\vspace{1.8cm}
{\large D.\ Fioravanti, F.\ Ravanini and M.\ Stanishkov\footnote{
Address after June 1, 1995:
I.N.R.N.E. - Sofia, Bulgaria\\
E-mail: fioravanti@bo.infn.it, stanishkov@bo.infn.it,
ravanini@bo.infn.it}
}\\
\vspace{.5cm}
{\em I.N.F.N. Sect. and Dept. of Physics - Univ. di Bologna\\
     Via Irnerio 46, I-40126 BOLOGNA, Italy}
\end{center}
\vspace{1cm}

\renewcommand{\thefootnote}{\arabic{footnote}}
\setcounter{footnote}{0}

\begin{abstract}
We propose an alternative description of 2 dimensional Conformal Field
Theory in terms of Quantum Inverse Scattering. It is based on the
generalized KdV systems attached to $A_2^{(2)}$, yielding the
classical limit of Virasoro as Poisson bracket structure.
The corresponding T-system is shown to coincide with the one
recently proposed by Kuniba and Suzuki. We
classify the primary operators of the minimal models that commute with
all the Integrals of Motion, and that are therefore candidates to
perturb the model by keeping the conservation laws. For our
$A_2^{(2)}$ structure these happen to be
$\phi_{1,2},\phi_{2,1},\phi_{1,5}$, in contrast to the
$A_1^{(1)}$ case, studied by Bazhanov, Lukyanov and
Zamolodchikov~\cite{BLZ},
related to $\phi_{1,3}$.
\end{abstract}
\newpage

{\bf 1}.
The great success of two-dimensional Conformal Field Theory (CFT) in the last
years is mainly due to its large symmetry. It possesses in fact an infinite
number of conserved charges closing a kind of generalization of $W_{\infty}$
algebra. Actually, it was shown~\cite{BPZ} that even the Virasoro subalgebra
generated by the higher momenta of the stress-energy tensor
is often enough for classifying the fields present in the theory and for the
computation of their 4-point correlation functions. Perturbation
of CFT with some relevant operator leads the system out of the scale
invariant fixed (or critical)
point. For specific perturbations, an infinite number of commuting conserved
charges survives~\cite{Z},
thus leading to an integrable theory. The
classification of all the possible integrable perturbations of a given CFT
is an important open question. At present, little is known about the
integrable field theories themselves. One of the main reasons for this is that,
despite the presence of an
infinite number of conserved charges, these generate an
abelian algebra from which it is difficult to extract
information about
correlation functions and physical quantities in general.
It is well known, however, that the integrable
field theories describe effectively a factorized scattering theory (FST)
of (massive~\cite{ZZ}
or massless~\cite{ZZ2}) particles. The on-shell information, including the
asymptotic states created by the so-called Zamolodchikov-Faddeev (ZF) operators
$Z_a(\theta)$ and the factorized S-matrices, is available at present for a
large class of such FST. Understanding the relation between the two
descriptions of the integrable field theory and in particular between the
ZF operators and the corresponding operators of the underlying Quantum
Field Theory is an important open problem.

On the other hand, CFT itself is an integrable theory -- it can be shown in
fact that the large symmetry algebra mentioned above has (at least
one) infinite abelian subalgebra. A possible strategy, put forward recently
by Bazhanov, Lukyanov and Zamolodchikov~\cite{BLZ} consists in trying
to understand the aforementioned link between the particle description and
the field theory one by implementing a Quantum Inverse Scattering
Method (QISM) for CFT. Of course, if the interest is CFT
alone, this investigation seems quite academic, as there are simpler
and more efficient ways to solve it. However,
this approach may become very fruitful if one thinks to ``prepare'' the CFT
to be perturbed off-criticality in an integrable direction. Indeed,
off-criticality the Virasoro symmetry is lost, and the QISM results in the
most hopeful method today available to compute physical quantities
of the theory.
The proposal of~\cite{BLZ} is then first to map the CFT data into a QISM
structure {\em at criticality} and later study how to leave the critical
point by suitable ``perturbation'' of this structure. Only the first part
of this project is considered in~\cite{BLZ} and in the present paper.
\vskip .3cm

{\bf 2}.
The starting point of~\cite{BLZ}
is the classical limit ($c\to -\infty$) of CFT given
by the KdV system~\cite{gervais,KM}, which corresponds to $A_1^{(1)}$
in the Drinfeld-Sokolov~\cite{DS} classification\footnote{We recall
  that in~\cite{DS} a generalized {\em modified} KdV (mKdV) hierarchy
  is attached to each affine Lie Algebra $\cG$. Various Miura
  transformations relate it to the generalized KdV hierarchies, each
  one classified by the choice of a node $c_m$ of the Dynkin diagram
  of $\cG$. Nodes symmetrical under some automorphism of
  the Dynkin diagram lead, however, to the same hierarchy. The
  classical Poisson structure of this hierarchy is a classical
  $w(\tilde{\cG})$-algebra, where $\tilde{\cG}$ is the Lie algebra
  obtained by deleting the $c_m$ node.}, and describes the
isospectrality condition of the second order differential operator
\eq
L= \de^2 + U(u) - \lambda^2
\label{due}
\en
The classical limit is realized as $T(u)\to -\frac{c}{6}U(u)$,
$T(u)$ being the stress-energy tensor on a periodic
strip ($T(u+2\pi)=T(u)$) and the Virasoro algebra is turned into the
Poisson bracket algebra
\eq
\{U(u),U(v)\}=2(U(u)+U(v))\delta'(u-v)+\delta'''(u-v)
\label{due-bis}
\en
Eq.(\ref{due}) has a first order matrix
representation
\eq
{\cal L}=\de - \phi'(u) h - (e_0 + e_1)
\label{tre}
\en
where the $A_1^{(1)}$ generators $e_0,e_1,h$ in the
fundamental representation and canonical gradation read
\eq
e_0=\left(\begin{array}{cc} 0 & \lambda \\
                            0 & 0 \end{array}\right)\virg
e_1=\left(\begin{array}{cc} 0 & 0 \\
                            \lambda & 0 \end{array}\right)\virg
h  =\left(\begin{array}{cc} 1 & 0 \\
                            0 & -1 \end{array}\right)
\en
$\phi(u)$ is related to $U(u)$ by the Miura
transformation $U(u)=-\phi'(u)^2 -\phi''(u)$.

Of primary interest is the monodromy matrix $\bM_j(\lambda)$ of the operator
(\ref{tre}), computed in a generic irreducible representation $\pi_j$,
$j=0,\half,1,\sfrac{3}{2},...$ and
especially its trace $\bT_j(\lambda)$, which is known to be
a generating function for the
commuting integrals of motion. The quantization procedure~\cite{FL} consists
essentially in using the quantum deformations $(A_1^{(1)})_{q_{\pm}}$
instead of $A_1^{(1)}$, where
\eq
q_{\pm}=
e^{i\pi\beta_{\pm}^2} \virg
\beta_{\pm}=\sqrt{\frac{1-c}{24}} \pm \sqrt{\frac{25-c}{24}}
\virg \beta_+=\frac{1}{\beta_-}
\en
and a free scalar field
\eq
\phi(u)=Q + Pu + i\sum_{n \not= 0} \frac{a_n}{n} e^{inu}
\label{cinque}
\en
\eq
[Q,P]=i\sfrac{\beta_{\pm}^2}{2} \virg
[a_n,a_m]=\sfrac{\beta_{\pm}^2}{2}n\delta_{n,-m}
\en
The Miura transformation translates, at the quantum level,
into the celebrated Feigin-Fuchs construction of the CFT through the
screened free boson (\ref{cinque})
\eq
-\beta_{\pm}^2 T(u) =
:\phi'(u)^2: + (1-\beta_{\pm}^2) \phi''(u) + \frac{\beta_{\pm}^2}{24}
\en
The two possible choices of parametrising with $\beta_{\pm}$
correspond to the two $(A_1)_{q_\pm}$ structures known to be present
in $M_{p,p'}$, namely $q_-=e^{i\pi\frac{p}{p'}}$ and
$q_+=e^{i\pi\frac{p'}{p}}$~\cite{Sierra}. In~\cite{BLZ} only the
$\beta_-$ parametrisation was considered. From $c=1-6(\beta_+ +
\beta_-)$ one can see that the classical limit $c\to -\infty$ can be
equally well realized by sending $\beta_+\to\infty$, $\beta_-\to 0$
(the choice in~\cite{BLZ}), or vice-versa.

The quantum version of the trace of the monodromy matrix
$\bT_j(\lambda)$ happens to be an entire function of $\lambda^2$, the
coefficients in the expansion being given by the ``non-local IM''. It
exhibits an essential singularity at infinity. The corresponding asymptotic
expansion generates the well known ``local IM'' of CFT. One can compute
perturbatively the first few terms in the expansion and check that these
$\bT$-operators (as well as their eigenvalues) satisfy a functional relation
\eq
\bT_j(q_{\pm}^{1/2}\lambda)\bT_j(q_{\pm}^{-1/2}\lambda) =
1+\bT_{j-\frac{1}{2}}(\lambda) \bT_{j+\frac{1}{2}}(\lambda)
\label{sette}
\en
When $\beta_{\pm}^2$ is a rational number ($\frac{p}{p'}$ or
$\frac{p'}{p}$), $q_{\pm}$ are roots of unity and quantum group
truncation takes effect, reducing the system (\ref{sette}) to a finite one.
The main conjecture of~\cite{BLZ} is that the states in the Hilbert
space of the $M_{p,p'}$ minimal model are in one to one correspondence
with the solutions of the corresponding truncated T-system having
suitable analytic properties (see~\cite{BLZ} for details). Thus, the
T-system describes completely the chiral Hilbert space in the particle
description.

The two objects $\int_0^{2\pi}du:e^{\pm 2\phi}:$ both commute with the local
IM~\cite{SY,EY,KM}.
One of them can be chosen to be, in the Feigin-Fuchs construction
language, the screening charge. There are two possible choices for
screening charges, which
amount to choose one of the two parametrisations with
$\beta_{\pm}$. The $\beta_-$ choice of~\cite{BLZ} identifies
the other vertex operator with the $\phi_{1,3}$ one, as a simple
calculation of conformal dimensions shows~\cite{EY}. Perturbations by
an operator whose charge commutes with a list of IM, means that, although
modified, the whole series of IM continues to be conserved
off-criticality. In other words, the structure (\ref{sette}) should be
kept by the perturbation, although the analytic properties of the
solutions should change drastically. The $\beta_-$ choice then can be
seen as the most efficient QISM description of
the particle structure ``ready'' to describe
perturbation by the $\phi_{1,3}$ relevant operator. It is interesting
to note that the other choice ($\beta_+$) leads instead to the
``perturbing'' operator $\phi_{3,1}$. This latter is never
relevant. However, it is known that this operator is ``formally''
integrable and has the important role of attracting towards infrared
the integrable fluxes $M_{p,p'} \to M_{2p-p',p}$. The application of
this observation to massless integrable theories has yet to be
explored.
We only mention here that if we apply the truncation of
(\ref{sette}) as explained in sect. 5
to this situation, we get a Y-system {\em formally}
identical to the one describing
the flux $M_{p+1,p+2}\to M_{p,p+1}$.
\vskip .3cm

{\bf 3}. In what follows we give an example of the non-uniqueness of
this description of CFT as integrable theory. Indeed, consider the
generalized KdV equations corresponding to the two vertices $c_0$ and $c_1$
of the Dynkin diagram of $A_2^{(2)}$ in the Drinfeld-Sokolov~\cite{DS}
classification
\bea
c_0 &:& \partial_t U = \de^5 U + 5 U \de^2 U +5 \de U \de^2 U + 5 U^2 \de U
\nonumber \\
c_1 &:& \partial_t U = \de^5 U + 10U\de^3 U + 25 \de U \de^2 U + 20 U^2 \de U
\label{dieci}
\eea
As the usual KdV, both equations (\ref{dieci}) are Hamiltonian.
Their second Hamiltonian structures are associated with the
Hamiltonians

\eq
H^{(0)} = 3(\de U)^2 -16 U^3 \virg H^{(1)}= 3 (\de U)^2 - U^3
\en
Here and in the following, the superscript in parenthesis $^{(0)}$ and
$^{(1)}$ refer to the $c_0$ and $c_1$ cases respectively.
The crucial observation is that the Poisson bracket algebra of the fields
$U(u)$ corresponding to these two second hamiltonian structures
coincides again
with the classical ($c\to -\infty$) limit of the Virasoro algebra,
eq.(\ref{due-bis}). The systems
(\ref{dieci}) describe isospectral deformations of {\em third}
order differential operators
\eq
L^{(0)} = \de^3 + U \de + \de U - \lambda^3 \virg
L^{(1)} = \de^3 + U \de - \lambda^3
\label{tredici}
\en
Eqs. (\ref{dieci}) can be obtained directly by
{\em reduction} of the Boussinesq equation, which describes the classical
limit of CFT having extended $W_3$-algebra symmetry~\cite{KM}. There are two
consistent reductions of Boussinesq equation: $W=\de U$ and $W=0$,
leading to the first and second equation of (\ref{dieci}) respectively.
However this
observation is valid only at the classical level, since $(A_2^{(2)})_q$,
which is relevant for the quantum case, is an essentially nonlinear deformation
of $A_2^{(2)}$, and not just a twist of $(A_2^{(1)})_q$.
Being integrable, the equations (\ref{dieci})
possess an infinite number of conserved IM $I_s^{(i)}$, $i=0,1$ having
spin $s=1,5 \bmod 6$~\cite{frenkel}
One can compute them using the Lax pair
representations of (\ref{dieci}) and show that the Poisson
bracket algebra they close is abelian $\{I_k^{(i)},I_l^{(i)}\}=0$, $i=0,1$.
These IM should obviously be the classical limit
of the corresponding
quantum conserved charges of CFT, and indeed they happen to coincide with
the classical limit of the quantum IM written in~\cite{KM} for the
Boussinesq system, once the reductions $W=\de U$ (for $c_0$) and $W=0$
(for $c_1$) are enforced.

Let us consider the first order matrix realization of (\ref{tredici})
\eq
\cL=\de - \phi'(u)h - (e_0+e_1)
\label{quattordici}
\en
Written in the canonical gradation of $A_2^{(2)}$ (see appendix B
of~\cite{DS}) eq.(\ref{quattordici})
defines the Lax representation for the generalized {\em modified} KdV
(mKdV) corresponding to algebra $A_2^{(2)}$.
Now $h,e_0,e_1$ are the Cartan-Chevalley
generators of $A_2^{(2)}$ level 0 algebra
\eq
e_0 = \left( \begin{array}{ccc} 0 & 0 & \lambda \\
                                0 & 0 & 0 \\
                                0 & 0 & 0 \end{array} \right) \virg
e_1 = \left( \begin{array}{ccc} 0 & 0 & 0 \\
                                \lambda & 0 & 0 \\
                                0 & \lambda & 0 \end{array} \right) \virg
h   = \left( \begin{array}{ccc} 1 & 0 & 0 \\
                                0 & 0 & 0 \\
                                0 & 0 &-1 \end{array} \right)
\label{trentuno}
\en
By choosing instead to
represent the $h,e_0,e_1$ matrices in the two possible standard
gradations ($c_0$ or $c_1$), one obtains that the first component of
eq.(\ref{quattordici}) satisfies the first and second of
(\ref{tredici}) respectively.

The expressions (\ref{tredici}) are obtained if one takes $h,e_0,e_1$ in the
fundamental representation. One can however give meaning to (\ref{quattordici})
for general representations of $A_2^{(2)}$.
The irreducible representations $\pi_s$ relevant here are labelled by
an integer $s\geq 0$. From the solution to the equation
$\cL\Psi(u)=0$ the monodromy matrix can be easily written
\eq
\bM_s(\lambda) = \pi_s \left\{ e^{2\pi ikh} \cP \exp \lambda \int_0^{2\pi} du
(e^{-2\phi(u)}e_0 + e^{\phi(u)}e_1) \right\}
\en
(recall that the field $\phi$ has the quasi-periodicity
$\phi(u+2\pi)=\phi(u)+2\pi k$).
Its ``improved'' form
$\bL_s(\lambda) = \pi_s(e^{-i\pi k h}) \bM_s(\lambda)$
satisfies the Poisson bracket algebra
\eq
\{\bL_s(\lambda) \stackrel{\otimes}{,} \bL_{s'}(\mu)\}=
[r_{s,s'}(\lambda\mu^{-1}),\bL_s(\lambda)\otimes \bL_{s'}(\mu)]
\label{diciannove}
\en
where $r_{s,s'}$ is the classical r-matrix associated with $A_2^{(2)}$
\cite{Jimbo}.
It follows from (\ref{diciannove}) that the trace of the monodromy matrix
$\bT_s(\lambda)=\mbox{Tr} \bM_s(\lambda)$ closes an abelian Poisson
bracket algebra
\eq
\{\bT_s(\lambda),\bT_{s'}(\mu)\}=0
\en
One can check that this $\bT$-operator in the fundamental representation
($\bT_1$) is
indeed the generating function of the infinite number of classical IM
of the $A_2^{(2)}$ mKdV.

\vskip .3cm

{\bf 4}.
Let us turn now to the quantum case. Following~\cite{FL,BLZ} we define
the quantum monodromy matrix and the $\bL$-operator as follows
\eq
\bL_s(\lambda)=\pi_s\left\{e^{i\pi Ph}{\cal P}\exp \lambda \int du
(:e^{-2\phi}: q^h e_0 + :e^{\phi}: q^{-h/2} e_1)\right\}
\label{ventuno}
\en
\eq
\bM_s(\lambda)=\pi_s(e^{i\pi Ph})\bL_s(\lambda)
\en
where $\phi(u)$ is a free massless scalar field like (\ref{cinque}),
and $e_0,e_1,h$
are now Cartan-Chevalley generators of the affine quantum algebra
$(A_2^{(2)})_q$ for $q=e^{i\pi\beta^2}$:
\[
[e_i,f_j]=\delta_{ij}[h_j] \virg [h_i,e_j]=a_{ij}e_j \virg [h_i,f_j]=-a_{ij}f_j
\virg i,j=0,1
\]
\eq
h=h_0=-2h_1 \virg a_{00}=a_{11}=2 \virg a_{01}=-4 \virg
a_{10}=-1
\en
where $[a]=\frac{q^a-q^{-a}}{q-q^{-1}}$.
We shall comment later on the relation between
$\beta$ and $c$ in this case.
Similarly to the classical case we can give meaning to
(\ref{ventuno}) in any
irreducible representation of $(A_2^{(2)})_q$.

We briefly describe here these representations,
as, up to our knowledge, they are nowhere written explicitly
in the mathematical
literature. Denote the basic vector of the representation $\pi_s$ as
$|j,m\rangle$, $j=0,\frac{1}{2},1,...,\frac{s}{2}$, $m=-j,-j+1,...,j$.
We define the action of the generators of $(A_2^{(2)})_q$ on this basis by
\bea
h|j,m\rangle &=& 2m |j,m\rangle \nonumber \\
e_0|j,m\rangle &=& \sqrt{[j-m][j+m+1]}|j,m+1\rangle \nonumber \\
f_0|j,m\rangle &=& \sqrt{[j+m][j-m+1]}|j,m-1\rangle \label{ventiquattro}\\
e_1|j,m\rangle &=& \sqrt{e(j)[j-m+1]}|j+\half,m-\half\rangle +
                   \sqrt{e(j-\half)[j+m]}|j-\half,m-\half
                   \rangle \nonumber \\
f_1|j,m\rangle &=& \sqrt{e(j)[j+m+1]}|j+\half,m+\half\rangle +
                   \sqrt{e(j-\half)[j-m]}|j-\half,m+\half
                   \rangle \nonumber
\eea
and
\eq
e(j)=\frac{[j+1][j+\frac{1}{2}]}{[\frac{1}{2}][2j+2][2j+1]}
\{[\sfrac{s}{2}+1] + [\sfrac{s}{2}+\half] - [j+1] - [j+\half]\}
\en
is the solution of the recursive equation
\eq
\frac{[2j]}{[j]}e(j-\half) - \frac{[2j+2]}{[j+1]}e(j) = 1 \virg
e(\sfrac{s}{2})=0
\label{venticinque}
\en
One can verify by direct calculation that the definition (\ref{ventiquattro})
indeed ensures the closing of the $(A_2^{(2)})_q$ algebra provided
(\ref{venticinque}) is satisfied.

Let us now return to the operator (\ref{ventuno}). It can be shown that
$\bL_s(\lambda)$ so constructed satisfies the quantum Yang-Baxter equation
\eq
\bR_{ss'}(\lambda\mu^{-1})(\bL_s(\lambda)\otimes\buno)
(\buno\otimes\bL_{s'}(\mu)) = (\buno\otimes\bL_{s'}(\mu))
(\bL_s(\lambda)\otimes\buno) \bR_{ss'}(\lambda\mu^{-1})
\label{ventisei}
\en
where now $\bR_{ss'}$ is the quantum
$\bR$-matrix associated with $(A_2^{(2)})_q$.
For instance, for the product
of fundamental representations $\pi_1\otimes\pi_1$ one has a $9\times 9$
matrix given in~\cite{Jimbo}.

The definition (\ref{ventuno}) is understood
in terms of power series expansion in $\lambda$
\eq
\bL_s(\lambda)=\pi_s\left\{e^{i\pi Ph} \sum_{k=0}^{\infty} \lambda^k
\int_{2\pi\geq u_1 \geq ... \geq u_k \geq 0} du_1 ... du_k K(u_1) ... K(u_k)
\right\}
\label{ventotto}
\en
where
\eq
K(u)=:e^{-2\phi(u)}:q^h e_0 + :e^{\phi(u)}: q^{-h/2} e_1
\en
Similarly to the case considered in~\cite{BLZ}
an estimate of the singularity properties of the integrands shows
that the integrals in (\ref{ventotto})
should be convergent for $\beta^2<\half$
and need regularization for $\beta^2\geq\half$. The analytic properties of the
eigenvalues of $\bT_j$
are strongly influenced by this regularization. However, the general structure
of the recurrence relations (\ref{trentasei}), as well as
(\ref{sette}), having its roots in purely algebraic properties of
affine quantum algebras, is expected to
remain intact for all the $c<1$ CFT's.

A direct consequence of (\ref{ventisei}) is that
the trace of the quantum monodromy matrix
\eq
\bT_s(\lambda) \equiv \mbox{\rm Tr} \bM_s(\lambda)
\label{trenta}
\en
defines a commuting operator $[\bT_s(\lambda),\bT_{s'}(\mu)]=0$ which
is the generator of quantum local and non-local IM.
In the case of the fundamental representation
$\pi_1$, one easily computes $\bT_1(\lambda)$ in terms of power
series expansion around $\lambda=0$
\eq
\bT_1(\lambda) = 2 \cos 2\pi P + \sum_{n=1}^{\infty} \lambda^{3n} Q_n
\label{trentadue}
\en
where
\bea
Q_n &=& q^{3n/2} \int_{2\pi\geq u_1 \geq ... \geq u_{3n} \geq 0}
du_1 ... du_{3n}~\times \label{trentatre}\\
&\times&\left\{e^{2i\pi P} :e^{-2\phi(u_1)}::e^{\phi(u_2)}::e^{\phi(u_3)}: ...
:e^{-2\phi(u_{3n-2})}::e^{\phi(u_{3n-1})}::e^{\phi(u_{3n})}:\right.
\nonumber\\
&+& \left. e^{-2i\pi P} :e^{\phi(u_1)}::e^{\phi(u_2)}::e^{-2\phi(u_3)}: ...
:e^{\phi(u_{3n-2})}::e^{\phi(u_{3n-1})}::e^{-2\phi(u_{3n})}: \right\}
\nonumber
\eea
are the non-local IM. As a consequence of (\ref{trentadue}),
$\bT_1(\lambda)$ is
an entire function of $\lambda^3$. One can show that it also exhibits an
essential singularity at infinity. The analysis of the corresponding
asymptotic expansion should involve a hard Bethe Ansatz calculation
that we do not perform here. Even the ``rough estimate''~\cite{lesage}
proposed in~\cite{BLZ} for the behaviour of the leading term of the
asymptotic expansion becomes much more involved in the present case,
as the relevant terms in the expansion of $\bT_s(\lambda)$ do not
enjoy the same nice symmetries as in~\cite{BLZ}.
Our expectation, of course,
is that the coefficients in this expansion should be given
by the quantum version of the local IM~\cite{KM}. We intend to return
on this important problem in the near future~\cite{35}.

In the general case, eq.(\ref{trenta}) can be computed using the so-called
$\bR$-fusion procedure~\cite{KRS}.
Here we give only the first non-trivial terms
in the $\lambda$-expansion
\eq
\bT_s(\lambda)=\frac{\sin\frac{s+2}{2}x \sin\frac{s+1}{2}x}
{\sin x \sin\frac{x}{2}} + \lambda^3 A_s(x,a) Q_1 + O(\lambda^6)
\label{trentaquattro}
\en
where $x=2\pi P$, $a=\pi \beta^2$ and
\bea
A_s(x,a)&=&\sum_{l=0}^s \frac{1}{8\sin x \sin a \sin\sfrac{a}{4}}
\left[\frac{\sin (x-a)(l+1)}{\sin(x-a)}-
  \frac{\sin(x+a)(l+1)}{\sin(x+a)}\right] \nonumber \\
&\times& \frac{\cos\sfrac{a}{2}\sin\sfrac{a}{2}(s+\sfrac{3}{2})-
      \cos\sfrac{a}{2}\sin\sfrac{a}{2}(l+1)}
     {\cos\sfrac{a}{2}(l+1)\cos\sfrac{a}{2}l}
\label{trentacinque}
\eea
One can show, using the explicit form (\ref{trentaquattro}) that
$\bT_s(\lambda)$ satisfies
(at least to order $\lambda^3$) the fundamental
relation
\eq
\bT_s(q^{1/6}\lambda) \bT_s(q^{-1/6}\lambda) =
\bT_{s+1}(\lambda) \bT_{s-1}(\lambda)+
\bT_s(q^{1/3\beta^2}\lambda)
\label{trentasei}
\en
The very nice result is that this equation coincides with the one
recently conjectured in a completely different fashion in~\cite{Kun},
and already present essentially in~\cite{Resh}. Conversely, it is
interesting to note that, assuming (\ref{trentasei}) as correct and
expanding in $\lambda$, one gets, for each order in the expansion,
new curious identities. Their meaning and possible relations with
fermionic representations of characters have still to be clarified.

The possible choices of $\beta$ in the $(A_2^{(2)})_q$ case are
dictated by adapting the classical limit of the $A_1$ Feigin-Fuchs
construction to the two possible choices of Miura transformations,
labeled by $c_0$ and $c_1$. Moreover, the classical limit can be
realized in two ways, sending $\beta_+\to\infty$ or
$\beta_-\to\infty$. This gives 4 possibilities in total. One can see
that both operators $\int_0^{2\pi}du:e^{-2\phi}:$ and
$\int_0^{2\pi}du:e^{\phi}:$ commute with the IM~\cite{KM}. Following
the same reasoning as in the $A_1^{(1)}$ case, if one
chooses $:e^{-2\phi}:$ as screening operator, then $:e^{\phi}:$ is the
perturbing field. The two parametrizations with $\beta_{\pm}$
correspond to the two possible choices for the screening operator and
give $:e^{\phi}:=\phi_{1,2}$ and $:e^{\phi}:=\phi_{2,1}$ respectively.
However, one is also free to choose $:e^{\phi}:$ as screening
operator. This leads to the identification of $:e^{-2\phi}:$ with
$\phi_{1,5}$ for $\beta_-$ and $\phi_{5,1}$ for $\beta_+$. The
$\phi_{5,1}$ operator is never relevant, and the same considerations
as for the $\phi_{3,1}$ operator apply here. It would be very
interesting to understand which fluxes are attracted by this
integrable direction in the $M_{p,p'}$ models. The $\phi_{1,5}$
operator is relevant only
for $p<p'/2$, $p'<6$~\cite{Vays}, but can play important roles in different
contexts, see e.g.~\cite{SRT}. It is a sort of ``dual'' operator to
$\phi_{2,1}$, in the same sense as $\phi_{3,1}$ is ``dual'' to
$\phi_{1,3}$.
\vskip .3cm

{\bf 5}.
In general (\ref{sette}) and (\ref{trentasei}) could be considered as
recursive relations for $\bT_s(\lambda)$. For $q$ root of 1 however
the quantum group truncation operates and (\ref{sette},\ref{trentasei})
become closed systems of functional equations. This important fact
allows the authors of~\cite{BLZ} to do a crucial conjecture: the
solutions of (\ref{sette}) (and we feel to extend the same argument to
(\ref{trentasei}) of course) having the suitable asymptotic behavior
and analytic properties (see~\cite{BLZ} for details) are the whole set of
eigenvalues $t_s(\lambda)$ of $\bT_s(\lambda)$ in the Hilbert space of the
model. Therefore systems (\ref{sette}) and (\ref{trentasei}) are
complete descriptions of the (chiral) Hilbert space of $c<1$ RCFT's, i.e.
minimal models.
In~\cite{BLZ} a transformation is given to pass from (\ref{sette}) to
a form which makes contact with scattering theory, the so-called
(scale-invariant limit of)
Y-system appearing in the context of Thermodynamic Bethe Ansatz (TBA)
equations. We have to signal that the success of this transformation is
however limited to the case of $M_{2,2n+3}$ models examined
in~\cite{BLZ}.
Comparison with the recently published~\cite{Tateo}
TBA system for $\phi_{1,3}$ perturbations of $M_{p,p'}$ other than
$M_{2,2n+3}$ shows
that in general the relation between the $\bT$-system and the Y-system
of TBA should be governed by a more complicate relation than the one
proposed in~\cite{BLZ}. This point out to the existence of a non-trivial
structure (sort of generalized Yangian?) relating the Virasoro
irreducible representations building up the chiral Hilbert space of
the $M_{p,p'}$ model and the organization of the same space under
subspaces diagonalising the $\bT_j$'s. Its understanding
passes through the decomposition of Virasoro structures into
$\bT_j$-diagonal ones (or vice-versa), a problem on which light can be
shed by use of identities relating bosonic (Virasoro) and fermionic
($\bT_j$-diagonal) forms of characters~\cite{Berk-McCoy}.
Also, this approach points out that the searched
particle description could be related to the spectrum generating {\em spinon
bases} recently introduced in RCFT~\cite{spinon}. Work in the
direction of clarifying these issues is in progress~\cite{35}, in this
short letter we do not deal with this problem
completely. Nevertheless, we would like to mention two cases where the
Y-system can be {\em formally} obtained from the T-system with a simple
transformation. The meaning of these and more complicated
transformations that we intend to report elsewhere still have to be
clarified.
\begin{itemize}
\item from the system (\ref{sette}), choosing
  $Y_j(\theta)=(t_{j+\half}(\lambda)t_{j-\half}(\lambda))^{-1}$ and
  $\lambda=e^{\frac{\xi}{\xi+1}\theta}$ for $j=\half,1,...,n$ and
  $Y_0=Y_{n+\half}=0$ one recovers the well known Y-system for {\em
    unitary} minimal models perturbed by $\phi_{1,3}$~\cite{Al3}.
\item A similar situation works in the case of system
  (\ref{trentasei}). Take
  $Y_s(\theta)=t_s(q^{1/3\beta^2}\lambda)(t_{s-1}(\lambda)
  t_{s+1}(\lambda))^{-1}$ and $\lambda=e^{\frac{\xi+1}{\xi}\theta}$,
  $\xi=\frac{p}{p'-p}$. With the $\beta_+$ parametrization of the choice
  $:e^{-2\phi}:$ as screening operator this gives the $T_1\diamond A_k$
  systems~\cite{RTV}, recently studied in~\cite{SRT} in connection with the
  $M_{p,2p-1}$ models perturbed by $\phi_{2,1}$. The same system, with
  $\lambda=e^{\frac{\xi}{\xi+1}\theta}$ and choice $\beta_-$ for the
  $:e^{\phi}$ screening, gives the other half of the perturbation
  studied in~\cite{SRT}, namely $M_{p,2p+1}$ perturbed by
  $\phi_{1,5}$. The phase shift appearing on the r.h.s. of
  (\ref{trentasei}) luckily happens to be readsorbed in these two
  cases thanks to the periodicity properties of the solutions of the Y-system.
\end{itemize}
All the other cases involve much more subtle transformations taking
into account the delicate structure of the quantum group
truncations. In particular, in the $A_2^{(2)}$ case, this latter is
even more involved and not yet completely clarified. This is the
reason why it is not possible here to find an analog of the
``most simple truncation'' reported in~\cite{BLZ} for
$A_1^{(1)}$. Even the simple models $M_{2,2n+3}$ perturbed by
$\phi_{1,2}$ happen to have a Y-system that can be obtained from
(\ref{trentasei}) only after a quite involved transformation.
\vskip .3cm

{\bf 6}.
To summarize, by exhausting the possible Drinfeld-Sokolov reductions
leading to classical Virasoro as Poisson algebra of their second
hamiltonian structure, we have found six integrable operators
($\phi_{1,3},\phi_{3,1},\phi_{1,2},\phi_{2,1},\phi_{1,5},\phi_{5,1}$),
which seem to exhaust the {\em generically} integrable primary
operators (i.e. integrable for {\em all} minimal models).
Of course there are other integrable perturbations for {\em specific}
minimal models. All those known so far seem however to be explained or
by some coincidence of operators,
or because the model happens to enjoy a larger symmetry
than the pure Virasoro one, and the operator is a generically
integrable one for this larger $W$-symmetry.

More generally, this
approach could give a simple method to classify all the generically
integrable perturbations of coset models of type
$\tilde{\cG}_k\times\tilde{\cG}_1 / \tilde{\cG}_{k+1}$ (complications
could arise for non-simply-laced $\tilde{\cG}$) in a way similar to
the observation recently made by Vaysburd in ref.~\cite{Vays}. The
connection with that paper could even go further, giving possibly some
key of understanding of the duality observed there between the affine
Toda system related to specific perturbed CFT and the affine quantum
symmetry of the model.

Of course many problems are still unsolved in this new and promising
approach initiated in~\cite{BLZ}. We mention the already signaled
problem to clarify in general the transformation from the T to the
Y-system, thus linking this approach with the recent developments
of~\cite{Tateo}.
Another (related) development is the link with fermionic
representation of characters~\cite{Berk-McCoy} that
could reveal for example the presence of
eventual spectrum generating dynamical symmetries, in the form of {\em
spinon bases} and shed new
light on the structure of correlation functions of the
fields of the model, in a way suitable for off-critical extensions.

The completeness of such program is related to extending
the above constructions out of the conformal point by ``perturbing'' the
integrable structures described here and in~\cite{BLZ}.
The results of~\cite{Luk} could help to do some progress in this direction.
If this ``perturbation'' could be
not too difficult to imagine (although quite hard to implement) for {\em
massive} theories, it creates new problems for {\em massless} ones.
The particle structure of the conformal point in this case is not related to
the ultraviolet theory, but to the infrared one, thus creating the new problem
of perturbing the infrared point by an irrelevant (but integrable!) operator,
a very delicate problem indeed~\cite{Berk}. The possibility to define
QISM structures related to irrelevant integrable operators could be of
help in this case. We hope to
return to all these intriguing issues elsewhere~\cite{35}.
\vskip .3cm

{\bf Acknowledgments} - We thank F. Lesage, J.O. Madsen, G. Sotkov
and I. Vaysburd for useful discussions. M.S. acknowledges I.N.F.N.-Bologna
for the kind hospitality and financial support during part of this work. The
same does F.R. with ENSLAPP-Annecy. This work was supported in part by
NATO Grant CRG 950751.


\begin{thebibliography}{99}
\bibitem{BLZ}
V.V.Bazhanov, S.L.Lukyanov and A.B.Zamolodchikov,
preprint CLNS 94/1316 -- RU-94-98 -- hep-th/9412229
\bibitem{BPZ}
A.A.Belavin, A.M.Polyakov and A.B.Zamolodchikov, \NP{B241} (1984) 333
\bibitem{Z}
A.B.Zamolodchikov, Adv. Stud. Pure Math. {\bf 19} (1989) 641
\bibitem{ZZ}
A.B.Zamolodchikov and Al.B.Zamolodchikov, Ann. Phys {\bf 120} (1979) 253
\bibitem{ZZ2}
A.B.Zamolodchikov and Al.B.Zamolodchikov, \NP{B379} (1992) 602
\bibitem{gervais}
J.L.Gervais, \PL{B160} (1985) 277 and 279
\bibitem{KM}
B.A.Kupershmidt and P.Mathieu, \PL{B227} (1989) 245
\bibitem{DS}
V.G.Drinfeld and V.V.Sokolov, J.Sov.Math. {\bf 30} (1984) 1975
\bibitem{FL}
V.A.Fateev and S.Lukyanov, \IJMP{A7} (1992) 853 and 1325
\bibitem{Sierra}
C.Gomez and G.Sierra, \NP{B352} (1991) 791
\bibitem{SY}
R.Sasaki and I.Yamanaka, Adv. Stud. Pure Math. {\bf 16} (1988) 271
\bibitem{EY}
T.Eguchi and S.-K.Yang, \PL{B224} (1989) 373
\bibitem{frenkel}
B.Feigin and E.Frenkel, \PL{B276} (1992) 79; hep-th/9310022 to appear
in Lect. Notes in Math., vol.{\bf 1620}
\bibitem{Jimbo}
M.Jimbo, \CMP{102} (1986) 537
\bibitem{lesage}
P.Fendley, F.Lesage and H.Saleur, preprint USC-94-16 hep-th/9409176
\bibitem{35}
D.Fioravanti, M.Stanishkov and F.Ravanini, work in progress
\bibitem{KRS}
P.P.Kulish, N.Yu.Reshetikhin, E.K.Sklyanin, Lett. Math. Phys. {\bf 5}
(1981) 393
\bibitem{Kun}
A.Kuniba, J.Suzuki, J. Phys. {\bf A28} (1995) 711
\bibitem{Resh}
N.Yu.Reshetikhin, Lett. Math. Phys. {\bf 7} (1983) 205
\bibitem{Vays}
I.Vaysburd, \PL{B335} (1994) 161
\bibitem{SRT}
M.Stanishkov, F.Ravanini and R.Tateo, preprint hep-th/9411085,
\IJMP{A} to appear
\bibitem{Tateo}
R.Tateo, \PL{B355} (1995) 157
\bibitem{Berk-McCoy}
A.Berkovich and B.McCoy, preprint BONN-TH-94-28 hep-th/9412030
\bibitem{spinon}
P.Bouwknegt, A.W.W.Ludwig and K.Schoutens, preprint hep-th/9504074 and
references therein
\bibitem{Al3}
Al.B.Zamolodchikov, \NP{B358} (1991) 497 and 524
\bibitem{RTV}
F.Ravanini, R.Tateo and A.Valleriani, \IJMP{A8} (1993) 1707
\bibitem{Luk}
S.L.Lukyanov and Y.Pugai, preprint RU-94-41 hep-th/9412128
\bibitem{Berk}
A.Berkovich, \NP{B356} (1991) 655
\end{thebibliography}
\end{document}